\documentclass[nofootinbib,aps,preprint,tightenlines,amssymb,letterpaper]{revtex4}

\renewcommand{\a}{\alpha}

\renewcommand{\d}{\delta}\newcommand{\D}{\Delta}
\newcommand{\e}{\epsilon}

\renewcommand{\L}{\Lambda}
\renewcommand{\t}{\theta}

\renewcommand{\o}{\omega}
\renewcommand{\O}{\Omega}

\newcommand{\p}{\phi}

\renewcommand{\u}{\mu}

\newcommand{\cI}{{\mathcal I}}

\newcommand{\cO}{{\mathcal O}}

\newcommand{\real}{\mathbb{R}}

\newcommand{\SU}{\mathrm{SU}}

\newcommand{\tr}{\mathrm{tr}}

\newcommand{\dr}{\partial}

\newcommand{\w}{\wedge}

\renewcommand{\i}{\imath}

\newcommand{\be}{\begin{equation}}
\newcommand{\ee}{\end{equation}}
\newcommand{\beq}{\begin{eqnarray}}
\newcommand{\eeq}{\end{eqnarray}}

\begin{document}
\begin{titlepage}
\title{\large Diffeomorphisms and spin foam models}
\author{ Laurent Freidel}
\email{lfreidel@perimeterinstitute.ca}
\affiliation{\vspace{2mm}Perimeter Institute for Theoretical
Physics\\ 35 King street North, Waterloo  N2J-2G9,Ontario,
Canada\\} \affiliation{Laboratoire de Physique, \'Ecole Normale
Sup{\'e}rieure de Lyon \\ 46 all{\'e}e d'Italie, 69364 Lyon Cedex
07, France }

\author{David Louapre}
\email{dlouapre@ens-lyon.fr} \affiliation{\vspace{2mm}Laboratoire
de Physique, \'Ecole Normale Sup{\'e}rieure de Lyon \\ 46
all{\'e}e d'Italie, 69364 Lyon Cedex 07, France}\thanks{UMR 5672
du CNRS}

\date{\today}

\begin{abstract}
We study the action of diffeomorphisms on spin foam models. We
prove that in 3 dimensions, there is a residual action of the
diffeomorphisms that explains the naive divergences of state sum
models. We present the gauge fixing of this symmetry and show that
it explains the original renormalization of Ponzano-Regge model.
We discuss the implication this action of diffeomorphisms has on
higher dimensional spin foam models and especially the finite
ones.
\end{abstract}
\maketitle
\end{titlepage}

\section{Introduction}

Spin foam models are an attempt to describe the geometry of
spacetime at the quantum level. They give a construction of
transition amplitudes between initial and final spatial geometries
labeled by spin network states. A spin foam is a 2-dimensional
cell complex $F$ with polygonal faces labeled by representations
$j_f$ and edges labeled by intertwining operators $i_e$. Given a
spin foam we associate an amplitude  $A_f, A_e$ and $A_v$ to the
faces, edges and vertices of the spin foam respectively. They
depend only locally on the spins $j_f$ and intertwiners $\i_e$,
e-g the vertex amplitude is computed using the spins labeling the
faces incident to that vertex and intertwiners on incident edges.
 The
amplitude of a spin foam is then computed as the product of these
local weights :
\begin{equation}
     Z_F(j_f,i_e) = \prod_{f \in F_2} A_{f}(j)
            \prod_{e \in F_1} A_{e}(j,\i)
            \prod_{v \in F_0} A_{v}(j,\i).
\label{sfa}
\end{equation}
The corresponding partition function  associated to the two
dimensional cell complex $F$ is given by summation over all
admissible spins and intertwiners
\begin{equation}
    Z(F) = \sum_{\{j_f\},\{\i_e\}} Z(j_f,\i_e).
\label{partition.function}
\end{equation}
Spin foams can be understood to be dual to triangulations. The
vertex of a spin foam is dual to a codimension $0$ simplex. We can
therefore think about the amplitude $A_v$ as the `bulk' term,
interpreted as corresponding to the discrete version of the
amplitude $e^{iS}$. The edges and faces amplitudes are dual to
codimension $1$ and $2$ simplices and we will refer to them as the
discrete {\it measure} of integration. There is so far a general
consensus on the form of the vertex amplitudes, in 3d they are
given by the 6-j symbol \cite{PR} and by the 10-j symbol in 4d
\cite{BC}.

However, there are some open debates on the possible measures one
should use in the state sum. A measure for 4d euclidean gravity
was first proposed in \cite{Depietrietal} by DePietri et al. using
group field theory technics. This measure is such that the
summation (\ref{partition.function}) diverges if we do not have a
positive cosmological constant. Another proposal was made later in
\cite{perez} by Perez and Rovelli and it was shown that this
measure leads to convergent amplitudes \cite{perezalone}.
Numerical studies of these models and some others were made by
Baez et al. in \cite{BaezC}, we refer the reader to this paper for
a more detail account of this issue. It is a key issue since the
infinite versus finite models have very different properties. Let
us remark that the divergence of the partition function is not a
good reason to discard the corresponding models since we are
interested in correlation functions or transition amplitudes which
are ratio of divergent amplitudes and could be well defined. In
this paper we will address this issue first in the context of
three dimensional gravity.

In retrospect, the very first spin foam model was the
Ponzano--Regge model of 3-dimensional riemannian quantum gravity
\cite{PR}. In this model, the vertex amplitude is given by the 6-j
symbol, one of its key property is the fact that its
semi-classical asymptotics is governed by the exponential of the
Regge action. The measure was uniquely determined in order to
obtain a partition function invariant under refinement of the
triangulation. This choice of measure is such that the partition
function is divergent and it is only after a regularization and
the division by an ad hoc divergent factor sometimes called the
{\it anomaly} that the state sum is formally independent of the
choice of triangulation. In the original paper of Ponzano-Regge
the division by this infinite factor was required in order to have
a well defined continuum limit of the partition function. In all
subsequent papers on this model the same reason for the overall
factor was always advocated.

Our aim in this paper is to explain this divergence as the
infinite gauge volume of a remaining gauge symmetry. This symmetry
is the {\it translational} part of the local Poincar\'e symmetry
 and is classically equivalent to the diffeomorphism symmetry.
It is clear that the choice of a triangulation (or a spin foam)
breaks the full covariance of the theory, however this does not
mean that there is no residual action of the diffeomorphism group
on a fixed spin foam. We show in three dimensions that indeed
there is a residual action of the diffeomorphism group  which acts
at the vertices of the fixed triangulation. This result is in fact
known to specialist of Regge calculus \cite{Rocwill}. We also
prove that the infinite anomaly factor is necessary in order to
divide out the volume of this residual symmetry group.

The plan of the paper is as follows. In section
\ref{sec:classical}, we recall the classical gauge symmetries of
3d gravity and show how they are related to each other. In section
\ref{sec:discrete}, we present the construction of the
Ponzano-Regge model, using a discretization of the partition
function. We insist on the way the gauge symmetries are
implemented at this level. In particular we describe how to
implement the translational symmetry. In section
\ref{sec:divergences}, we relate the divergence of the
Ponzano-Regge model with the volume of the translational symmetry
and carry a gauge fixing procedure. In section
\ref{sec:discussion}, we conclude with a discussion on the general
implications of these results for higher dimensional models. We
argue that we can generally expect a residual action of
diffeomorphisms on spin foams, leading to a physical
interpretation of the divergences. We discuss  the consequences of
this observation for the finite spin foam models.
\section{Classical gauge symmetries in 3D gravity}\label{sec:classical}
In this part we recall the different gauge symmetries of the
classical action of 3d gravity, and how they are related together.
We consider the first order formalism for 3d gravity. The field
variables are the triad frame field $e_\mu^i$ ($i=1,2,3$) and the
spin connection $\o_\mu^{i}$. The metric is reconstructed as usual
from the triad $g_{\mu \nu}=e_\mu^i \eta_{ij} e_{\nu}^j$ where
$\eta = (+,+,+)$ for euclidean gravity and $\eta= (-,+,+)$ for
lorentzian gravity. In the following, we will denote by $e^i,
\o^i$ the one forms $e_\mu^idx^\mu, \o_\mu^i dx^\mu$. We also
introduce the $SU(2)$ Lie algebra generator $J_i$, taken to be
$-i/2$ times the Pauli matrices,  satisfying $[J_i,J_j] =
\epsilon_{ijk}\ \eta^{kl} J_l$, where $\e_{ijk}$ is the
antisymmetric tensor. The trace is such that
$\tr(J_iJ_j)=-\frac{1}{2}\delta_{ij}$. One can thus define the Lie
algebra valued one-forms $e=e^i J_i$ and $\o=\o^i J_i$. The action
is
\begin{equation}\label{eqn:BFaction}
S[e,\o]=-\frac{1}{2}\int_{M} \e_{ijk}\ e^i \w F^{jk}(\o) =
\int_{M} \tr(e \w F(\o)),
\end{equation}
where $\wedge$ is the antisymmetric product of forms  and
$F(\o)=d\o+ \o \wedge \o$ is the curvature of  $\o$. The equations
of motion of this theory are \beq
d_\o e =0, \label{eom1}\\
F(\o)=0 \label{eom2},\eeq where $d_\o =d +\o$ denotes the
covariant derivative.

Classically, this theory has three kind of symmetries. First, it
is invariant under local Lorentz gauge symmetry:
\begin{eqnarray}
\d^L_X \o &=& d_\o X, \nonumber \\
\d^L_X e &=& [e,X], \label{eqn:classLorentz}
\end{eqnarray}
parametrized by a Lie algebra element $X$. It is also invariant
under diffeomorphism, for a vector field $\xi^\u$, the action  is
\begin{eqnarray}
\d^D_\xi e=d(\i_\xi e)+\i_\xi(d e), \nonumber \\
\d^D_\xi \o = d(\i_\xi \o)+\i_\xi(d\o), \label{eqn:diffeoA}
\label{eqn:diffeoE}
\end{eqnarray}
where $\i_\xi$ denotes the interior product. These are the usual
gauge symmetries of gravity. However this theory admits another
symmetry that we call \textit{translational symmetry}, given by
\begin{eqnarray}\label{trans}
\d^T_\phi \o &=& 0, \nonumber \\ \d^T_\phi e &=& d_\o \phi,
\end{eqnarray}
for $\phi$ in the Lie algebra. As seen by inserting this
transformation in the action (\ref{eqn:BFaction}), this symmetry
is due to the Bianchi identity $d_\o F=0$. These three types of
symmetries are not all independent, one can see that
\begin{eqnarray}
\d^D_\xi e&=& \d^L_{(i_\xi \o)}e + \d^T_{(i_\xi e)}e + i_\xi(d_\o e),\\
\d^D_\xi \o&=& \d^L_{(i_\xi \o)}\o + \d^T_{(i_\xi e)}\o
+i_\xi(F(\o)).
\end{eqnarray}
If one uses the equations of motion (\ref{eom1}) and (\ref{eom2}),
one clearly sees that on-shell we have \beq \label{diffgauge}
\d^D_\xi = \d^L_{(i_\xi \o)} + \d^T_{(i_\xi e)}. \eeq This shows
that on-shell, the diffeomorphism symmetry is recovered as a
combination of Lorentz and translational symmetry, for field
dependent parameters of transformation $X=\i_\xi \o,\ \phi=\i_\xi
e$. Note that the combination $\d^L +\d^T$ is in fact a local
Poincar\'e gauge symmetry. Lets denote by $P_i$ the generators of
the translations, they commute between themselves and satisfy $
[J_i,P_j]=\e_{ijk}\eta^{kl} P_l$. We can introduce a Poincar\'e
connection $A= \o^i J_i + e^i P_i$. The symmetry $\d^L +\d^T$ is
just the local Poincar\'e gauge symmetry for this connection.

\section{Gauge symmetries in discrete quantum
gravity}\label{sec:discrete}

In this section we recall briefly the discretization procedure
leading to the Ponzano-Regge model for 3d euclidian quantum
gravity. We emphasize the implementation of gauge symmetries and
explain how the translational symmetry arises at the level of this
model. We are interested in the computation of the partition
function of 3 dimensional gravity, i-e formally \be\label{part} Z=
\int De D\o\ e^{iS(e,\o)}. \ee In order to do this computation we
will first choose a triangulation $\Delta$, discretize the action
with respect to this triangulation and then compute the
discretized path integral. The continuum limit is then obtained by
refining the triangulation. This methodology and the resulting
computation are not new, they have been considered and refined
several time in the literature \cite{Ooguri,Nielsen, KF}, our
approach will be very close to the treatment in \cite{KF}. The new
point of our approach is to insist on the fact that when  we
compute the partition function (\ref{part}), and more generally
transition amplitudes, one integrates over all $e$ and $A$ {\it
modulo} gauge transformations. In the continuum one usually uses
the Fadeev-Popov procedure. In the discrete approach what one
should do is identify the residual gauge symmetries that are left
after choosing the triangulation, and divide out the volume of
this residual discrete gauge symmetry. In the following we do the
proper analysis and find that there is a residual Lorentz gauge
symmetry acting at the vertices of the dual triangulation {\it and
} a residual translational symmetry associated with the vertices
of the triangulation.

\subsection{Discretized model}
The action (\ref{eqn:BFaction}) can be discretized in order to
formulate a discretized version of 3d quantum gravity. One
considers a triangulation $\D$ of the manifold where all the edges
are oriented. The triad $e$ is a 1-form and as such it is
naturally integrated on one dimensional structures. We integrate
it along the edges $e\in\D$ of the triangulation, and replace it
by the collection of Lie-algebra elements $X_e$ obtained in this
way. To discretize the connection field, one considers the dual
2-complex $\D^*$. One can naturally consider the holonomy of the
connection (which is a 1-form) along the edges $e^*$ of the dual
two complex. This assigns a group element $g_{e^*}$ to each dual
edge $e^*$. The discretized curvature is obtained as the holonomy
of the connection around a whole dual face $f^*$, i-e as the
ordered product of corresponding group elements living on the dual
edges
\begin{equation}\label{eqn:gface}
g_{f^*}=\overrightarrow{\prod}_{e^* \subset f^*} g_{e^*}.
\end{equation}
To be more precise, in order to define this holonomy we chose a
vertex of the dual triangulation $v^*$ and chose a path
$P_{v^*,f^*}$ in the dual triangulation which starts at $v^* $,
goes to a point on $f^*$, goes around the face and come back by
the same initial path. The path should be oriented in a way
compatible with the orientation of the edge $e$ dual to $f^*$.
Different choices of initial points $v^*$ or different choices of
pathes $P_{v^*,f^*}$ are equivalent, since the corresponding group
elements we obtain are related by transformations $ g_{f^*}\to g_0
g_{f^*}g_0^{-1}$, which are gauge transformations, as will be
proved in the following. If we take the logarithm of this group
element (\ref{eqn:gface}), we get a Lie algebra
element\footnote{We restrict the Lie algebra element $Z$ to be in
the region around $0$ in which $\exp$ is an isomorphism between
this region and the group.} $Z_{f^*}$ such that
$g_{f^*}=e^{Z_{f^*}}$. We denote this Lie algebra element by $Z_e$
for simplicity, this is valid since there is a one to one
correspondence between edges of $\D$ and dual faces. So within
this discretization the dynamical variables are $(X_e, g_{e^*})$,
and the action is simply expressed as
\begin{equation}\label{eqn:discreteaction}
S[X_e,g_{e^*}]=\sum_e \tr(X_e Z_e).
\end{equation}
The partition function is \be \label{eqn:partfun} Z(\D)=
\int_{\mathfrak{g}^E} \prod_e dX_e \int_{G^{E^*}}
\prod_{e^*}dg_{e^*}\ e^{i\sum_e \tr(X_e Z_e)}, \ee where
$\mathfrak{g}$ denotes the Lie algebra, $E$ and $E^*$ denote the
number of edges in $\D$ and $\D^*$. It is clear that this
partition function does not depend on the choice of the
orientation, since both $X_e$ and $Z_e$ change sign if we change
the orientation of $e$. Before calculating this discretized
partition function, we are going to show how the gauge symmetries
of the classical action are reflected in this discretization.

\subsection{Discrete gauge symmetries}
Since $g_{e^*} $ is the holonomy of the connection along the dual
edge $e^*$, the gauge group acts by left and right multiplication
at the vertices of the dual lattice
\begin{equation}
g_{e^*} \to k^{-1}_{s(e^*)}\ g_{e^*}\ k_{t(e^*)},
\end{equation}
where $s(e^*)$ (resp. $t(e^*)$) denotes the source (resp. the
target) of the edge $e^*$. Since $g_{f^*}$ is a product of edges
group elements (see eq.\ref{eqn:gface}) starting and ending at the
same vertex $v^*$, the gauge group acts by conjugation on this
variables. The discrete action (\ref{eqn:discreteaction}) is
invariant under the transformation
\begin{eqnarray}
Z_e &\to& g_{v^*}^{-1} Z_e g_{v^*}, \\
X_e &\to& g_{v^*}^{-1} X_e g_{v^*}.
\end{eqnarray}
This is the implementation of the classical local Lorentz
transformation (\ref{eqn:classLorentz}) at the level of the
discrete model.

The continuum translational transformation is
\begin{equation}\label{tsym}
 \d e = d\phi +[\o,\phi]
\end{equation}
where $\phi$ is a zero-form valued in the Lie algebra. We have
seen that $e$ is naturally integrated on the edges of the
triangulation, $X_e=\int_e e$. The 0-form $\phi$ is naturally
discretized at the vertices of the triangulation in terms of a
collection of Lie algebra elements $\Phi_v$. We therefore expect
the discrete transformation to be
\begin{equation}
\delta X_e= \Phi_{t(e)}-[\O^{t(e)}_e, \Phi_{t(e)}] - \Phi_{s(e)} +
[\O^{s(e)}_e, \Phi_{s(e)}].
\end{equation}
Recall that we choose an orientation of the edges, so it is clear
that the discretization of $d\Phi$ leads to
$\Phi_{t(e)}-\Phi_{s(e)}$. We define $\O_e^v$ to be an integrated
version of $\o$ on $e$ \textit{starting from} $v$
\begin{equation}
\O^v_e \sim \int_{v\stackrel{e}{\to}} \o +...
\end{equation}
in the limit where all edges lengths are small. This explains why
we have a plus sign for $[\O^{s(e)}_e, \Phi_{s(e)}]$ and a minus
sign for $[\O^{t(e)}_e, \Phi_{t(e)}]$ . One can isolate the action
at a vertex
\begin{equation}\label{eqn:disctrans}
\d X_e=\Phi_v-[\O^v_e, \Phi_v].
\end{equation}
The expression (\ref{eqn:disctrans}) corresponds to the integrated
version of the classical translational symmetry (\ref{trans}). The
problem in this expression is that the connection $\o$ has
originally been discretized by integrating on the {\it dual} edges
$e^*$, and there is therefore no natural discrete expression for
the integration $\O^v_e$ of $\o$ on the original edge $e$. To see
how to deal with this problem, we examine the Bianchi identity
which is at the origin of this symmetry, and show how this
identity holds at the discretized level.

Classically, the Bianchi identity is
\begin{equation}
dF+[\o,F]=0.
\end{equation}
This is an identity about 3-forms. Integrated on a 3d volume $V$,
this leads to
\begin{equation}\label{eqn:intbianchi}
0=\int_{\dr V} F+\int_V [\o,F]
\end{equation}
One can consider  a vertex of the original triangulation and the
surface of dual 2-faces surrounding it. If the triangulation is
regular, this surface $S$ has the topology of a 2-sphere and one
can consider the integrated Bianchi identity
(\ref{eqn:intbianchi}) on the interior $B$ of this 2-sphere.
Decomposing $B$ as the disjoint union of cones obtained by linking
each dual face to the central vertex, one can rewrite it
\begin{equation}
\int_{S} F + \int_B [\o,F]= \sum_{e\supset v} Z_e +\sum_{e\supset
v} [\O^v_e,Z_e]
\end{equation}
where $\O^v_e$ is an integrated version of $\o$ on the edge $e$
surrounded by the corresponding cone. Note that we also choose to
define $\O_e^v$ as the integration of $\o$ along $e$
\textit{starting from} $v$, which means along an edge outgoing
from the interior of $B$, which is necessary to integrate in this
way the Bianchi identity. From this argument we expect that the
element $\O^v_e$ appearing in the translational symmetry
(\ref{eqn:disctrans}) can be given by the understanding of the
discrete Bianchi identity. Our analysis is closed to the one
already given in \cite{Nielsen} by Kawamoto et al. In our
computation their statements are made precise by a careful
analysis of the origin of the discrete Bianchi identity.

To do so, we begin by observing that if one considers the dual
faces surrounding the vertex, it exists an order on the faces
$f_1...f_n$ and a collection of group elements $g_{f^*_i}$
representing their curvature such that\footnote{Remember that we
had some freedom to chose a vertex in the dual triangulation and a
path connecting this vertex to the dual face. We need to use this
freedom in order for the identity (\ref{eqn:prodg}) to be true. We
don't spell out the details of the construction as it is standard,
and more or less obvious from a drawing.}
\begin{equation}\label{eqn:prodg}
\overrightarrow{\prod_{i=1..n}} g_{f^*_i} = 1
\end{equation}
One can now take the logarithm of this expression
\begin{equation}\label{eqn:logprodg}
\ln\left(\overrightarrow{\prod_{e\supset v}} e^{Z_e}\right)=0
\end{equation}
In the case of an abelian group, this expression is just $\sum
Z_e=0$, which corresponds to the discretized version of the
abelian Bianchi identity $dF=0$. However in the general case of a
non-abelian group, taking the logarithm leads to a more
complicated result which has to be expressed using the
Baker-Campbell-Hausdorff formula. In the appendix \ref{app:BCH},
we show (see eq. (\ref{eqn:BCHomega})) that the logarithm
(\ref{eqn:logprodg}) can be rewritten as
\begin{equation}\label{bianchiBCH}
\sum_{e\supset v} \left(Z_e+[\O^v_e,Z_e]\right)=0,
\end{equation}
where $\O^v_e$ is a Lie algebra element explicitly given in terms
of the $Z_e$ on the edges meeting at the vertex $v$ \be \O^v_e =
\sum_{e'\supset v, e'\neq e} Z_e + \sum_{e',e''\supset v}
c_{ee'e''} [Z_{e'},Z_{e''}] + \cdots, \ee the dots stand for
higher commutator terms constructed in terms of $Z_e$ and
$c_{ee'e''}$ are explicit coefficients given in
(\ref{eqn:approxO}). See (\ref{eqn:exactO}) for the complete
formal expression of $\O_e^v$ including all higher commutators.
The expression (\ref{bianchiBCH}) is the Bianchi identity at the
discretized level, and its construction provides an expression for
the elements $\O^v_e$.

It is now clear that due to the discretized Bianchi identity, the
discretized action is invariant under the symmetry \beq
\delta X_e = \Phi_v-[\O^v_e,\Phi_v] \ \ \mbox{if}\ \ v \subset e \nonumber\\
\delta X_e =0 \ \ \mbox{if}\ \ v \not\subset e
\label{eqn:discretetopol} \eeq for $\Phi_v$ a Lie algebra element
associated to the vertex $v$. The variation of the discretized
action under this transform is
\begin{eqnarray}
\d S&=&\sum_{e\supset v} \tr(\Phi_vZ_e- [\O^v_e,\Phi_v]Z_e)\\
    &=&\tr\left[\Phi_v \left(\sum_{e\supset v} Z_e+[\O^v_e,Z_e]\right)\right] =0 \\
\end{eqnarray}
In the first equality we use cyclicity of the trace and the last
equality is due to the discretized Bianchi identity
(\ref{bianchiBCH}). The action is invariant under the transform
(\ref{eqn:discretetopol}) which is the discrete action of the
translational symmetry. We have only considered the action at one
vertex but this can be extended for all the vertices of the
triangulation, the transformation is parametrized by a Lie algebra
element $\Phi_v$ for each vertex.

\section{Gauge symmetries and divergences} \label{sec:divergences}

In this part we complete the construction of the Ponzano-Regge
model, taking into account the infinite volume of the
translational gauge symmetry.

\subsection{Division by the gauge volume}

We have seen that the local Lorentz gauge symmetry is
parameterized by group elements acting at the vertices of the dual
$\D^*$, while the translational symmetry is parameterized by Lie
algebra elements acting at the vertices of $\D$. If we denote by
$V$ (resp. $V^*$) the number of vertices of the triangulation
(resp. the dual complex), one has to divide the naive (non gauge
fixed) discretized partition function by the total gauge volume
$Vol(G)^{V^*}\times Vol(\mathfrak{g})^{V}$. $Vol(G)$ denotes the
volume of the Lorentz group, it is finite for euclidean
gravity\footnote{It is infinite in the  case of lorentzian gravity
$G=SL(2,R)$ but we can take care of it by appropriate
regularization, see \cite{3dLorentz}. In this paper we restrict
ourselves to euclidean gravity for simplicity, but considering the
lorentzian case will not really change our discussion.} $G=SU(2)$.
$Vol(\mathfrak{g})$ denotes the {\it infinite} volume of the Lie
algebra. Dividing by the gauge volume, the partition function
(\ref{eqn:partfun}) is rewritten
\begin{equation}\label{eqn:partfunfixed}
Z(\D)=\frac{1}{Vol(G)^{V^*}\times
Vol(\mathfrak{g})^{V}}\int_{G^{E^*}} \prod_{e^*} dg_{e^*}
\int_{\mathfrak{g}^{E}} \prod_e dX_e\ e^{i\sum_e \tr\left(X_e
Z_e\right)}.
\end{equation}
One can find a choice of measures on the Lie algebra and the group
\begin{eqnarray}
dX & = & \frac{1}{4\pi} r^2 dr\ \sin\p d\p\ d\psi \label{eqn:mesLiealg},\\
dg & = & \frac{1}{2\pi^2} \sin\t^2 d\t\ \sin\p d\p\ d\psi
\label{eqn:mesG}.
\end{eqnarray}
such that $V(G)=1$, and
\begin{equation}\label{delta}
\int_{\mathfrak{g}} dX e^{i\tr(XZ)}=\d(e^Z),
\end{equation}
where $\d$ is the delta function on the group for the measure
(\ref{eqn:mesG})\footnote{Strictly speaking, the equation
(\ref{delta}) is true only when $Z$ is around $0$.}. It is
expanded on the characters of the representations of $G$ using
Plancherel decomposition
\begin{equation}\label{plancherel}
\d(g)=\sum_j d_j \chi^j(g),
\end{equation}
where $\chi^j$ designs the character of the spin $j$
representation. The integrals over group elements in the partition
function (\ref{eqn:partfunfixed}) are performed using the
relations on the integrations of product of matrices of
representations. The resulting contributions are given by Wigner
$3j$-symbols on the dual edges, which recombine into $6j$ symbols
associated to tetrahedra. One thus get the Ponzano-Regge model
\begin{equation} \label{eqn:PRpart}
Z(\D)=\frac{1}{[Vol(\mathfrak{g})]^{V}} \sum_{\{j_e\}} \prod_e
d_{j_e} \prod_t \{6j\}.
\end{equation}
This expression was interpreted originally by Ponzano and Regge as
a partition function for discrete 3d gravity. In terms of
simplicial geometry, an edge carrying a spin $j$ is interpreted as
an edge of length $j+1/2$, from the fact that the asymptotic of
the $6j$-symbol leads to the Regge action for the tetrahedra with
lengths $j+1/2$, and the original observation due to Wigner that
the asymptotic of the square of the $6j$-symbol is interpreted as
the classical probability to have a tetrahedron with lengths
$j+1/2$. The equation (\ref{eqn:PRpart}) is purely formal since
both the spin state sum and the volume of the Lie algebra are
infinite. We therefore need to introduce a cut-off in order to
regularize this expression. Our prescription is to restrict the
Lie algebra elements to be $|X_e| < k$. This has for consequence
to restrict the summation in the state sum over a finite number of
spins. This can be seen from the Kirillov correspondence between
representations and coadjoint orbits in the Lie algebra. The
character of the spin $j$ representation is expressed by the
Kirillov formula
\begin{equation}
\chi^j(e^Z)={\int_{\cO_j} e^{i\tr(Z X)}\ d_j\vec{X} \over
\int_{\cO_{0}}  e^{i\tr(Z X)}\ d_{0}\vec{X} }
\end{equation}
where $\cO_j$ is the sphere of radius $2j+1$ and $d_j\vec{X}$ the
measure such that the volume of $\cO_j$ is equal to $2j+1$. This
shows that restricting the Lie algebra elements by $|X| < k$
restricts the representations to $2j +1 < k$ i-e $j\leq
\frac{k-2}{2}$. With this cut, the volume of the Lie algebra
becomes finite and equal to $V(k)=\frac{k^3}{3}$ with our choice
of measure (\ref{eqn:mesLiealg}). This can be also obtained from
the cut-off on the representations in the Plancherel formula
(\ref{plancherel}), since for large $k$
\begin{equation}
V(k)=\int_{|X|< k} dX e^{i\tr(X\cdot 0)} \sim \sum_{d_j < k} d_j^2
\sim \frac{k^3}{3}.
\end{equation}
The renormalized partition function is thus obtain by sending the
cut-off to infinity
\begin{equation}
Z(\D)=\lim_{k \to \infty} \sum_{\{d_{j_e}<k\}} \prod_v
\frac{1}{V(k)} \prod_e d_{j_e} \prod_t \{6j\}.
\end{equation}
The partition function for another triangulation $\D'$ obtained by
refinement of $\D$ is formally such that $Z(\D')=Z(\D)$. In the
original paper of Ponzano-Regge, the renormalization by
$V(k)^{-V}$  was motivated by the requirement to have invariance
under refinement. Our argumentation shows that this is not just an
ad-hoc renormalization, but it arises from the division of the
volume of the translational symmetry acting at the vertices.

\subsection{Gauge fixing procedure}
In this part, we carry a precise Fadeev-Popov gauge fixing
procedure for the translational symmetry. To do so, we choose a
maximal tree $T$ in the 1-skeleton of the triangulation $\D$. A
maximal tree is a graph touching every vertex of the
triangulation, without forming a closed loop. The main property of
a maximal tree is that it exists an unique path in $T$ between any
two vertices. A natural way to gauge fix the symmetry is to impose
that all $X_e$ on $T$ are zero. Since a maximal tree with $V$
vertices contains $V-1$ edges, this gauge fixing procedure will
fix all the symmetry except a global translation. We isolate one
of the vertices in the tree (called the root). This induces an
order in the tree and an orientation on the edges, each edge being
oriented along the direction of the path from the root to each
vertex. Each edge has thus a source $s(e)$ and a target $t(e)$.
Moreover each vertex $v$ can be unambiguously associated to the
edge $e$ such that $v=t(e)$. The (inverse of the) Fadeev-Popov
determinant for this procedure is
\begin{equation}
\D^{-1}=\int_{\mathfrak{g}^E} [\prod_e d\Phi_{t(e)}]\ \prod_{e\in
T}
\d(\Phi_{t(e)}-[\O^{t(e)}_e,\Phi_{t(e)}]-\Phi_{s(e)}+[\O^{s(e)}_e,\Phi_{s(e)}]).
\end{equation}
To compute the F-P determinant, we have to compute the jacobian of
the transformation, or equivalently the wedge product of all the
$d\tilde\Phi_{t(e)}$ where
\begin{equation}
\tilde\Phi_{t(e)} =
\Phi_{t(e)}-[\O^{t(e)}_e,\Phi_{t(e)}]-\Phi_{s(e)}+[\O^{s(e)}_e,\Phi_{s(e)}].
\end{equation}
One can see that the variable $\Phi_{t(e)}$ for an external edge
$e$ appears only in $\tilde\Phi_{t(e)}$, thus we have
\begin{equation}
d\tilde\Phi_{t(e)} =
d\left(\Phi_{t(e)}-[\O^{t(e)}_e,\Phi_{t(e)}]\right) + \cdots
\end{equation}
where the dots are for terms that will give zero when taking the
wedge product with others $d\tilde\Phi$, since this product does
not contain $d\Phi_{t(e)}$. This reasoning apply for all external
edges, and can be continued for other edges. In particular we have
\begin{equation}
d\left(\Phi_{t(e)}-[\O^{t(e)}_e,\Phi_{t(e)}]\right) =
(1+|\O^{t(e)}_e|^2) d\Phi_{t(e)}
\end{equation}
where $|\O^{t(e)}_e|^2$ denotes the square of the norm of
$\O^{t(e)}_e$ as Lie algebra element. Therefore one gets for the
F-P determinant
\begin{equation}
\D=\prod_{e\in T} (1+|\O^{t(e)}_e|^2)
\end{equation}
From now we denote $\O_e=\O_e^{t(e)}$. The partition function on
the triangulation $\D$ for a gauge fixing along the tree $T$
becomes
\begin{equation}
Z(\D,T)=\prod_{e\in \D/T} \int dX_e \prod_{e^*\in \D^*} \int
dg_{e^*} \left(\prod_{e\in T} (1+|\O_e|^2)\right)
e^{i\tr(\sum_{e\in \D/T} X_e Z_e)},
\end{equation}
where the $Z_e$ and $\O_e$ depend on the group elements. The
integration on the remaining $X_e$ variables can be performed,
leading to
\begin{equation}
Z(\D,T)=\prod_{e^*\in \D^*} \int dg_{e^*} \left(\prod_{e\in T}
(1+|\O_e|^2)\right) \prod_{e\in \D/T} \d(e^{Z_e})
\end{equation}
Now one can prove that if the $Z_e$ are zero on $\D/T$, as imposed
by the delta functions, then the $\O_e$ are zero on the maximal
tree $T$ and the F-P determinant reduces to 1. Consider an edge
$e$ in $T$, one of its vertices $v$ and the corresponding $\O_e$.
$\O_e$ is defined (see (\ref{eqn:exactO})) in terms of the $Z$
variables of all edges meeting at $v$. Suppose first that this
vertex is an external vertex of $T$, which means that $e$ is the
only edge of $T$ meeting it. All other edges $e'$ meeting it are
in $\D/T$ and the $\d$ functions impose $Z_{e'}=0$ for them. In
that case, the only non-zero variable in the explicit expression
(\ref{eqn:exactO}) of $\O_e$ is $Z_e$, it is thus clear that all
the commutators vanish. So, for this external vertices, $\O_e$ is
zero. Moreover if we consider the Bianchi identity
(\ref{bianchiBCH}) at this vertex, separating, in the sum, $e$
from the other edges meeting, it says
\begin{equation}
Z_e + [\O_e,Z_e] + \sum_{e'\neq e} (Z_{e'}+[\O_{e'},Z_{e'}])=0
\end{equation}
As $Z_{e'}=0$ and $\O_e=0$, this identity gives $Z_e=0$. Thus for
all external vertices and corresponding edges of $T$, $\O_e=0$ and
$Z_e=0$. If one removes all these edges from the tree, one is left
with new external vertices, for which the same reasoning apply.
The procedure can be extended to the whole tree and all the $\O_e$
and $Z_e$ are zero on the tree. The F-P determinant is thus one
and the partition function is rewritten
\begin{equation}
Z(\D,T)= \prod_{e^*\in \D^*} \int dg_{e^*} \prod_{e\in \D/T}
\d(e^{Z_e}).
\end{equation}
At the level of the discretized partition function, the gauge
fixing of the $X_e$ variables to zero for $e \in T$ is translated
into a projection on the spins $j_e=0$ for the edges of the tree
$T$
\begin{equation}
Z(\D,T)=\sum_{\{j_e\}} \left(\prod_{e\in T}
\d_{j_e,0}\right)\prod_e d_{j_e} \prod_t
\{6j\}.\label{eqn:partfunproj}
\end{equation}
In the triangulation, this has the following consequences : if we
set $j_e=0$ for an edge $e$ belonging to $n$ triangular faces,
each of these faces is suppressed, and the two other edges of each
face are identified. We call this move the collapse of the edge
$e$. One can see that this is a topological move, so if we perform
the collapse of $\D$ along any tree $T$ and we denote the
resulting triangulation $\D_T$, this triangulation is equivalent
to the original one. The relevance of this definition lies in the
fact that
\begin{equation}
Z(\D,T)=Z(\D_T).
\end{equation}
Note that $\D_T$ is well-defined\footnote{as a generalized
triangulation} since the tree structure prevents us to try to
collapse an edge previously collapsed due to the identification
arising in the process. Note also that it still remains a global
translational symmetry acting at the vertex which was the root of
the tree, and which has to be gauge fixed. Finally, the result is
formally independent of the choice of a maximal tree :
$Z(\D_T)=Z(\D_{T'})$. This is clear since on one hand different
choices of trees amounts to different choices of gauge and it is
usually expected that after gauge fixing the theory is BRST
invariant, expressing the fact that all the gauge fixings are
equivalent. On the other hand, we have seen that the collapsed
triangulations along different trees are topologically equivalent,
leading to the same conclusion.

\subsection{Positive cosmological constant case}
We consider now the case of gravity with positive cosmological
constant. The euclidian space with positive cosmological constant
$\L$ is a 3-sphere of radius $1/\sqrt{\L}$. This is a space of
finite volume $1/\L^{3/2}$ (computed for the normalized Haar
measure for $S^3$). In such a space, the maximum geodesic length
is obtained for the half-perimeter of a great circle, namely
$L_{max}=\pi/\sqrt{\L}$. Since a maximum length exists in such
space, we expect that to be translated at the level of the quantum
model as a cut-off in the allowed spins. Also in the discrete
models, the translational symmetry acts at the vertices by
translation in the spacetime. In a positive cosmological constant
space, the volume accessible by translation is finite and equal to
the volume of the space. As the space is of finite volume, it acts
as a cut-off for the translational symmetry, since it forbids
arbitrary large translations of the vertices. We therefore expect
a \textit{finite} gauge volume corresponding to the volume of the
space $1/\L^{3/2}$.

These two expectations concerning the cut-off on the lengths and
the finite gauge volume are precisely what happens in the
Turaev-Viro model\cite{TV}. The Turaev-Viro model is obtained as a
deformation of the Ponzano-Regge model using the quantum group
$\SU_q(2)$ instead of $\SU(2)$. It depends on an additional
parameter $k$ such that $q=e^{2i\pi/k}$. This parameter can be
linked in several ways with the cosmological constant $\L$. First
it is well-known that the spins of the representations of
$\SU(2)_q$ are strictly smaller than $\frac{k-1}{2}$. This is
clear since the quantum dimension of the spin $j$ representation
\begin{equation}
[d_j]_k=\frac{\sin\left(\pi(2j+1)/k\right)}{\sin(\pi/k)}
\end{equation}
is zero when $j=\frac{k-1}{2}$. We have seen that the length for
an edge carrying a spin $j$ is given by $j+1/2$, so from the point
of view of the Turaev-Viro model, the maximum length is
$L_{max}=\frac{k}{2}$. If we identify it with the maximum geodesic
length in the 3-sphere of radius $\frac{1}{\L}$ we get
$k=\frac{2\pi}{\sqrt{L}}$. This interpretation of the level $k$
can also be done using the link between Turaev-Viro for
$q=e^{2i\pi/k}$ and Chern-Simons theory at level $k$ \cite{KFvol}.
The same interpretation is also manifest in the asymptotic of the
quantum 6j-symbol \cite{Mizoguchi} which leads to the exponential
of the Regge action with cosmological constant $\L=4\pi^2/k^2$.

It is well known that the triangulation independence of the
Turaev-Viro model requires the multiplication of the state sum by
the factor $Vol(\SU(2)_q)^{-V}$ where
\begin{equation}
Vol(\SU(2)_q)=\sum_{j=1}^{(k-2)/2} [d_j]^2_k =
\frac{k}{2\sin^2(\pi/k)}
\end{equation}
For a small cosmological constant (large $k$), it behaves as
\begin{equation}
Vol(\SU_q(2)) \sim \frac{k^3}{2\pi^2}\sim \frac{4\pi}{\L^{3/2}}.
\end{equation}
which scales as the volume of the 3-sphere for cosmological
constant $\L$. This is not a surprise if we interpret
$Vol(SU(2)_q)$ as the volume of the translational group acting at
the vertex.

One can notice that despite the fact that the quantum dimension
$[2j+1]_k$ for large $k$ is asymptotic to the dimension of the
classical representation $2j+1$, the large $k$ behavior of the
gauge volume is different in the Turaev-Viro model and in the
Ponzano-Regge regularization
\begin{equation} \sum_{j=1}^{(k-2)/2} [d_j]^2_k
\not\sim \sum_{j=1}^{(k-2)/2} d_j^2
\end{equation}
This fact seems puzzling but can be understood geometrically. We
have seen that the LHS is interpreted as the volume of a 3-sphere
while the RHS is the volume of a 3-ball.

\section{Discussion}\label{sec:discussion}

In this paper we have proven that a carefully discretized spin
foam model for 3-dimensional gravity possess not only the usual
Lorentz gauge symmetry, but also a local translational symmetry,
acting at the vertices of the triangulation, and parameterized by
Lie algebra elements. At the classical level, this symmetry is
related to the diffeomorphism symmetry, so we can interpret this
result as the existence of a residual action of the diffeomorphism
group on 3-dimensional triangulated space-time.

We have shown that the volume of this symmetry is infinite and
scales with the number of vertices of the 3-dimensional
triangulation. If we correctly quantize the theory via path
integral we have to divide out this infinite gauge volume. The
prescription we obtain for the regularized partition function
after the division by the volume of the gauge group is the same as
the one proposed a long time ago by Ponzano and Regge. However, in
their paper and in {\it all} subsequent paper on the subject, the
infinite volume factor was required in order to implement the
continuum limit. Our result prove that the requirement of gauge
invariance is enough to explain this factor and the fact that the
partition function possess the correct continuum limit is a
consequence of this requirement without any further
renormalisation. We also show that gauge fixing this symmetry
amounts to compute the partition
 function on a collapsed triangulation.

So what does this example teach us on higher dimensional spin foam
models? We can think of spin foam models as discretized models of
gravity where the vertices of the spin foam are dual to higher
dimensional simplices, the edges of the spin foam are dual to
codimension 1 simplices and the faces of the spin foam are dual to
codimension 2 simplices. The representation labels $j$ are carried
by the codimension 2 simplices. They are therefore interpreted as
a length ($l\sim l_{p}j$) in 3 dimensions and an area ($A \sim
l_{p}^{2} j$) in 4 dimensions. The first lesson from 3 dimensions,
is that we can expect a non trivial residual action of the
diffeomorphisms on the spin foam. The corresponding remaining
diffeomorphisms will not change the connectivity of the spin foam
but will act on its representation labels. This is not a new idea,
it was advocated a long time ago in the context of Regge calculus
by Rocek and Williams \cite{Rocwill}. In the context of Regge
triangulation, the labels are carried by the edges of the
triangulation and they proved that around the flat solution there
is an action of the diffeomorphism at the vertices of the
triangulation. This can be easily understood as follows. Suppose
we have  a Regge triangulation of flat space time. To obtain such
a triangulation, one can first triangulate $\real^{4}$ and then
put as Regge labels on the edges the lengths of the edges measured
with the flat metric.
 Now by action of a diffeomorphism on
$\real^{4}$ we can translate one vertex of the triangulation
without moving the others. This will gives us an other Regge
triangulation which differs from the first one by a relabeling of
the edges surrounding the vertex. This triangulation described the
same piecewise linear geometry. This indicates that Regge
triangulation still carry a residual action of the diffeomorphism
group. It was proven by Rocek and Williams that such
transformations leave the Regge action invariant.

In the 4d case, the situation in spin foam models is different
from Regge calculus. First, the labels are carried by the faces of
the triangulation, instead of the edges. Second, if we interpret
spin foams as evolving spin networks as was first done in the
literature \cite{ReisRov}, there is no obvious relation with
piecewise linear geometry. However one can still make a link
between the discretized partition function and spin foam models
\cite{KF}, and try to interpret translational symmetry as in 3d,
in terms of variation of the spin labels. At the classical level,
one can consider the bivector field $B^{IJ}=e^I \w e^J$, where
$e^I$ is the frame field. This is a 2-form and its natural
discretization on a triangulation is on faces, and is given by
$B_f^{IJ}=E_1^{I}\w E_2^{J}$, where $E_1$ and $E_2$ are the
discretization of the frame field $e$ along two edges of the face
$f$. Now at the discretized level, the spin $j_f$ carried by the
face $f$ can be interpreted as the norm of the simple bivector
$B_f$. In 3d we interpreted the remaining diffeomorphism symmetry
as changing the lengths of the edges, keeping the geometry fixed.
We want to argue that a similar interpretation survive in 4d. It
has been shown by Horowitz \cite{Hor} that in the first order
formalism of gravity, we can trade off diffeomorphism for
translational symmetry in any dimension. Around flat space, this
translational symmetry acts only on the continuous frame field as
$\d e^I=d_\o \Phi^I$ where $\Phi^I$ is a 0-form field. The
resulting transformation on the bivector field $B^{IJ}=e^I \w e^J$
is $\d B^{IJ}=d_w\Psi^{IJ}$ where $\Psi^{IJ}=\Phi^{[I}e^{J]}$ is a
frame dependent parameter of transformation. This is a 1-form,
naturally discretized on the edges of the triangulation. One
therefore expect the translational symmetry to act on the
discretized $B$ field at edges of the triangulation. This leads to
an action of this transformation on the 3-cells of the spin foam.
This is consistent with the fact that the Bianchi identity
responsible for this symmetry is a 3-form discretized along the
3-cells of the spin foam. In the light of what happens in 3d, one
expects a divergence for each 3-cell of the spin foam if the
diffeomorphism symmetry is not broken and if the geometry we are
integrating contains flat space. That's exactly the type of
divergences which is occurring in some of the spin foam models
first proposed in \cite{Depietrietal} and analyzed in
\cite{perez}. This open the possibility to give a physical
interpretation of these divergences as coming from a residual
action of the diffeomorphisms. Of course in 4 dimensions this is
only a plausibility argument so far and it deserves more study.

This argumentation therefore raise the question of the meaning of
finite spin foam models. There are two different types of spin
foam model which have a convergent sum over spins. The first type
of model is obtained when we consider euclidean gravity with a
positive cosmological constant, in that case the sum over spin is
restricted to be finite (the spins cannot be bigger than the
cosmological constant in Planck units). There is still an action
of the diffeomorphism group in 3-dimensions, but since the
3-sphere has a finite volume the volume of this group is finite.
This is consistent with the interpretation that a positive
cosmological constant  suppress spacetimes with large volumes.
This is extensively used in the context of dynamical triangulation
for instance, where a positive cosmological constant is needed to
make the summation convergent. In 4 dimension the inclusion of a
positive cosmological constant in the euclidean theory will also
lead to convergent spin foam models. We can argue in the same way
that this doesn't mean that the diffeomorphism group do not act on
the spin foams but only that the volume of the symmetry group is
finite in that case. In this case we therefore have a physical
interpretation of the finiteness of the model. It is important to
note that adding a cosmological constant help us only if its
positive and if the spacetime is euclidean. If we consider
lorentzian gravity with any cosmological constant or euclidean
gravity with a negative cosmological constant we cannot expect any
convergence of the models since the volume of the corresponding
homogeneous spaces are all infinite in these cases and so will be
the volume of the residual translational group.

The second kind of convergent models were first considered in
\cite{perez} and correspond to a different choices of edge
amplitudes. This is interpreted as a different choice of measure
in spin foams, as we discussed in the introduction. Since we
expect an action of diffeomorphisms on spin foam we have to
understand why this action does not translate in the amplitude by
an infinite factor. In these models there is no parameter we can
vary which can be interpreted as a  cosmological constant so we
cannot really argue that the convergence possess a clear physical
interpretation. An interpretation of the finiteness of the
amplitude is that diffeomorphism symmetry is broken in these
model. This means that we expect these models to assign different
amplitudes to gauge equivalent solutions. This can be explicitly
checked in three dimensions where a modification of the measure of
spin foam models similar to the one done in 4-dimension can be
achieved \cite{nonpert}. One possible interpretation for that
could be that these models are in fact gauged fixed models and
should not carry a representation of the diffeomorphism group.
This is however unlikely, since a Fadeev-Popov gauge fixing
procedure generally involves a highly non-local determinant. The
amplitude of finite spin foam models are expressed as a product of
local weights (cf eq. \ref{sfa}) which cannot be interpreted as a
Fadeev-Popov determinant. In \cite{BaezC} numerical simulation
showed that the amplitude of the finite model are dominated by the
spins $0$ and $1/2$ and for all practical purpose the summation
over spins can be restricted these ones. In the spin foam model,
the amplitude associated with a spin foam where a given face is
carrying the spin $0$ is equal to the amplitude of the spin foam
where this faces as been collapsed. So we can interpret the
amplitude $A(S) = \sum_{j=0,1/2}A(j)$ of a given spin foam $S$ as
a sum over all spin foams included in $S$ \be A(S) =
\sum_{S^\prime \subset S} \a^{N_{0}} \beta^{N_{1}} \gamma^{N_{2}},
\ee where $N_{0},{N_{1}},{N_{2}}$ denotes the number of $0,1,2$
cells of the spin foams and $\a = A_{v}(j=1/2)$,
$\beta=A_{e}(j=1/2)$, $\gamma=A_{f}(j=1/2)$ in the notation of
(\ref{sfa}). As such, the finite spin foam models are similar to
dynamical triangulations models with a fixed edge lengths $j=1/2$.
In dynamical triangulations, $\beta=1$ and the parameters $\a,
\gamma$ can be freely tuned, they depend on the Newton coupling
constant, the cosmological constant and the scale of the lattice
spacing \cite{Ambjorn}. It is known that dynamical triangulations
models do not carry a representation of the diffeomorphisms.
However, in the limit where $S$ becomes infinite we can tune the
parameters of the theory to recover a continuum limit in some
region of phase space and restore in this limit the action of
diffeomorphisms.

So the finite spin foam models have features very similar to
dynamical triangulations models, namely they  break the action of
diffeomorphism. It does not go along the line of the spin foam
philosophy which is to keep most of the symmetry of the continuum
theory in the intermediate triangulation dependant levels. So far
this requirement has been focused on local Lorentz
 symmetry but we have argued in this paper that there is the possibility to even fulfill this requirement for the local
Poincar\'e symmetry or diffeomorphism. One reason for this
requirement is the hope that it would help to get a final theory
with the correct symmetries when we get rid of the triangulation
dependance. This can be of course disputed and one can argue that
the symmetry is restored in the coarse grain limit even if it is
not implemented at the microscopic level. There is no very good
reason yet to prefer one mechanism over the other. However, we
still think that finite models have undesirable features. First,
they behave much alike dynamical triangulation models which is not
bad by itself but then it seems hard to expect spin foam to do
better than what has been done in this context. The second point
is that in dynamical triangulations models the parameters are
freely tuned and this is important to restore diffeomorphism
symmetry in the coarse grain limit. This is not a feature of spin
foam models so far.

In order to conclude, we have showed in this paper that there is a
residual action of translational symmetry on an individual spin
foam and that this explains the infinity or `anomaly' of the
Ponzano-Regge model. We have argued that such a residual action of
translational symmetry should be expected for higher dimensional
models. This would give a physical meaning to the divergences
observed in the 4D model proposed in \cite{Depietrietal}. On the
contrary we have disputed the relevance of the finite spin foam
models proposed in \cite{perez}. Our argumentation is by no mean
conclusive since it contains hypothesis and unresolved issues but
we hope that it will launch a fruitful and successful debate on
this issues.

\vspace{1cm}

\noindent\textbf{Acknowledgments :} We thank A. Ashtekar for
relevant questions on the gauge fixing procedure. We thank H.
Pfeiffer and A. Perez for discussions. D. L. is supported by a
MENRT grant and Eurodoc program from R\'egion Rh\^one-Alpes. L. F.
is supported by CNRS and an ACI-Blanche grant.

\appendix
\section{Baker-Campbell-Hausdorff formula}\label{app:BCH}
In this section we recall how to formally obtain the
Baker-Campbell-Hausdorff formula. In particular we obtain the
expression (\ref{bianchiBCH}) used in the text. We are interested
in the computation of the expression
\begin{equation}
X=\ln\left(e^{Z_1}e^{Z_2}...e^{Z_N}\right)
\end{equation}
for $Z_1,...Z_N$ Lie algebra elements. One can expand the
exponentials
\begin{equation}
X=\ln\left[\left(\sum_{p_1}\frac{Z_1^{p_1}}{p_1!}\right)\left(\sum_{p_2}\frac{Z_2^{p_2}}{p_2!}\right)...
\left(\sum_{p_N}\frac{Z_N^{p_N}}{p_N!}\right)\right]
\end{equation}
Isolating the term $p_1=p_2=...=p_N=0$ one can rewrite it
\begin{equation}\label{eqn:BCH}
X=\ln\left[1+\sum_{\stackrel{p_1...p_N}{p_1+...+p_N\geq
1}}\frac{Z_1^{p_1}}{p_1!}\frac{Z_2^{p_2}}{p_2!}...\frac{Z_N^{p_N}}{p_N!}\right]
\end{equation}
Expanding now the logarithm one gets
\begin{equation}
X=\sum_{k=1}^{+\infty} \frac{(-1)^{k-1}}{k}
\left(\sum_{\stackrel{p_1...p_N}{p_1+...+p_N\geq
1}}\frac{Z_1^{p_1}}{p_1!}\frac{Z_2^{p_2}}{p_2!}...\frac{Z_N^{p_N}}{p_N!}\right)^k
\end{equation}
Expanding the term to the power $k$, one gets a sum over all the
ways to take the orderer product of $k$ terms in the sum, namely
\begin{equation}\label{eqn:expandBCH}
X=\sum_{k=1}^{+\infty} \frac{(-1)^{k-1}}{k} \sum_{\cI^k}
\left[\frac{Z_1^{p^1_1}}{p^1_1!}\frac{Z_2^{p^1_2}}{p^1_2!}...\frac{Z_N^{p^1_N}}{p^1_N!}\right]
\left[\frac{Z_1^{p^2_1}}{p^2_1!}\frac{Z_2^{p^2_2}}{p^2_2!}...\frac{Z_N^{p^2_N}}{p^2_N!}\right]
...
\left[\frac{Z_1^{p^k_1}}{p^k_1!}\frac{Z_2^{p^k_2}}{p^k_2!}...\frac{Z_N^{p^k_N}}{p^k_N!}\right]
\end{equation}
where the sum is over the set $\cI^k$
\begin{equation}
\cI^k=\left\{p_i^j \in \mathbb{N}, i=1\cdots N, j=1\cdots k |
\sum_i p_i^j \geq 1 \ \forall j \right\}
\end{equation}
One thus obtains an expression of $X$ as a polynome in the $Z_i$.
Each monome is obtained as a product of $k$ subterms, the
condition $\sum_i p_i^j \geq 1\ \ \forall j$ states that each of
these subterms is non-empty. $X$ is a Lie algebra element, while
each monome in the sum at the LHS is generically an element of the
universal enveloping algebra $U(\mathfrak{g})$. The Lie algebra
element of $U(\mathfrak{g})$ are characterized by the fact one can
give $U(\mathfrak{g})$ a coproduct such that the Lie algebra
elements are the primitive elements of this coproduct. One can
apply at both sides of the equation a projector $\pi :
U(\mathfrak{g}) \to \mathfrak{g}$ which projects on the Lie
algebra elements of $U(\mathfrak{g})$. It exists different
projectors of this type, we will use the most common namely the
Dynkin projector. It is defined in the following way, consider a
monome which is written as $x_1 x_2...x_M$ for
$x_i\in\mathfrak{g}$ the projector is defined
\begin{equation}
\pi(x_1...x_M) = \frac{1}{M}
[[[...[[x_1,x_2],x_3]...]x_{M-1}],x_M]
\end{equation}
In particular, we have for $M \geq 2$
\begin{equation}\label{eqn:projrec}
\pi(x_1...x_M) = \frac{M-1}{M}[\pi(x_1...x_{M-1}),x_M]
\end{equation}
This is this projector which allows to express the
Baker-Campbell-Hausdorff formula in terms of commutators of the
originals $Z_I$. Consider one of the monome, i-e fix $k$ and a set
of indices $p_i^j$. We consider the action of this projector
\begin{equation}
\pi\left[\left(Z_1^{p_1^1}Z_2^{p_2^1}...Z_N^{p_N^1}\right)
\left(Z_1^{p_1^2}Z_2^{p_2^2}...Z_N^{p_N^2}\right) ...
\left(Z_1^{p_1^k}Z_2^{p_2^k}...Z_N^{p_N^k}\right)\right]
\end{equation}
The size of the monome is denoted $M=\sum_{i,j} p_i^j$. First, if
the monome is of size $M=1$, this means that there is only one
subterm (recall the subterms are asked to be non-empty), hence
$k=1$, and the projector acts as the identity. Now we consider the
general case $M\geq 2$. To use the expression (\ref{eqn:projrec}),
we need to understand what is the last term of the monome. We know
that the $k$-th subterm is non-empty since $\sum_i p_i^k \geq 1$.
We pick up the last non-zero element in the sequence
$p^k_1...p^k_N$ and denote it $p^k_{i_0}$. This means that the
monome ends with the element $Z_{i_0}$ and that the action of
$\pi$ on it is expressed as a commutator
\begin{equation}
\left[\frac{M-1}{M}
\pi\left(Z_1^{p_1^1}...Z_N^{p_N^1}...Z_1^{p_1^k}...Z_{i_0-1}^{p_{i_0-1}^k}
Z_{i_0}^{p_{i_0}^k-1}\right),Z_{i_0}\right]
\end{equation}
The total projection of all the monomes of size $M\geq 2$ of
(\ref{eqn:expandBCH}) can be rewritten as a sum of commutators of
a Lie algebra element with a single $Z_{i_0}$
\begin{equation}
\sum_{i_0} [\O_{i_0},Z_{i_0}]
\end{equation}
where $\O_{i_0}$ is defined by
\begin{equation}\label{eqn:exactO}
\O_{i_0}=\sum_{k=1}^{+\infty} \frac{(-1)^{k-1}}{k}
\sum_{\cI^k_{i_0}} \left(\prod_{i,j}\frac{1}{p_i^j!}\right)
\left(\frac{\sum_{i,j}p_i^j}{\sum_{i,j}p_i^j\ +1}\right)\ \
\pi\left[Z_1^{p^1_1}Z_2^{p^1_2}...Z_N^{p^1_N}
Z_1^{p^2_1}Z_2^{p^2_2}...Z_N^{p^2_N}...
Z_1^{p^k_1}Z_2^{p^k_2}...Z_{i_0}^{p^k_{i_0}}\right]
\end{equation}
where the set of indices $\cI^k_{i_0}$ is
\begin{eqnarray}
&& p_i^j\in \mathbb{N} | \sum_i p_i^j \geq 1, i=1..N, j=1..k-1\\
&& p_i^k\in \mathbb{N} | \sum_i p_i^k \geq 0, i=1..i_0
\end{eqnarray}
This set $\cI^k_{i_0}$ means that we have taken the set $\cI^k$
but cut the indices in the last subterm $k$ at order $i_0$ and
drop the condition $\sum p_i \geq 1$. Putting together the results
for $M=1$ and $M\geq 2$ we get
\begin{equation}\label{eqn:BCHomega}
X=\ln\left(e^{Z_1}...e^{Z_N}\right) = \sum_{i_0}
Z_{i_0}+[\O_{i_0},Z_{i_0}]
\end{equation}

One can estimate the first terms in the complicated expression of
the $\O_k$
\begin{eqnarray}
\O_k&=& \frac{1}{2}\sum_{k<i} Z_i \nonumber \\
    &+& \frac{1}{6}\sum_{i<j<k} [Z_i,Z_j]
-\frac{1}{6}\sum_{k<i<j} [Z_i,Z_j]+\frac{1}{12}\sum_{i}
[Z_i,Z_k] \nonumber \\
&+& ...\label{eqn:approxO}
\end{eqnarray}

\end{document}